\documentclass[preprint]{vgtc}               % preprint
\ifpdf%                                % if we use pdflatex
  \pdfoutput=1\relax                   % create PDFs from pdfLaTeX
  \usepackage{graphicx}                % allow us to embed graphics files
  \DeclareGraphicsExtensions{.pdf,.png,.jpg,.jpeg} % for pdflatex we expect .pdf, .png, or .jpg files
\else%                                 % else we use pure latex
  \usepackage{graphicx}                % allow us to embed graphics files
  \DeclareGraphicsExtensions{.eps}     % for pure latex we expect eps files
\fi%

\graphicspath{{figures/}{pictures/}{images/}{./}} % where to search for the images

\usepackage{microtype}                 % use micro-typography (slightly more compact, better to read)
\PassOptionsToPackage{warn}{textcomp}  % to address font issues with \textrightarrow
\usepackage{textcomp}                  % use better special symbols
\usepackage{mathptmx}                  % use matching math font
\usepackage{times}                     % we use Times as the main font
\usepackage{cite}                      % needed to automatically sort the references
\usepackage{tabu}                      % only used for the table example
\usepackage{booktabs}                  % only used for the table example
\usepackage{enumitem}
\usepackage{wrapfig}
\usepackage{hyperref}

\vgtccategory{Research}

\usepackage[caption=false, font=footnotesize]{subfig}

\vgtcinsertpkg

\title{Pilaster: A Collection of Citation Metadata Extracted From Publications on Visualization for the Digital Humanities}

\author{Alejandro Benito-Santos\thanks{e-mail: abenito@usal.es} %
\and Roberto Ther\'{o}n\thanks{e-mail: theron@usal.es}}
\affiliation{\scriptsize VisUSAL Research Group. Universidad de Salamanca, Spain}

\teaser{
  \includegraphics[width=\linewidth]{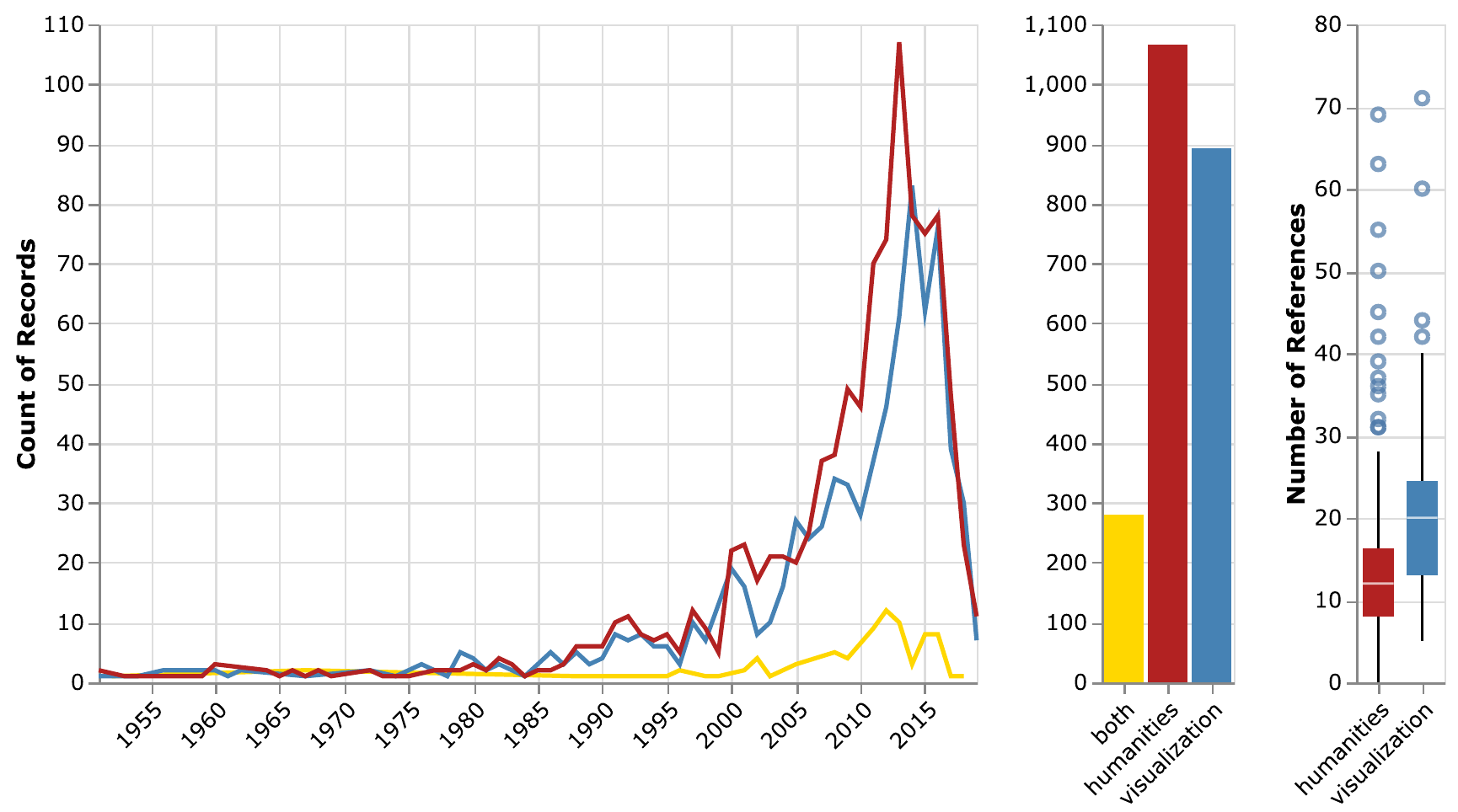}
  \caption{Three visualizations depicting different features of the citations dataset. From left to right, the first chart shows a portion (1950-2019) of the temporal distribution of resources in the citations dataset originating in the VIS community (blue), in the digital humanities community (red), or in both (yellow). The second chart shows total record counts in the citations dataset for each of these three categories. Finally, the whisker plots on the right display the distribution of reference list lengths by publication venue for records in the seed dataset.}
  \label{fig:temp-source-whiskers}
}

\abstract{
In this paper, we present Pilaster (\url{https://visusal.github.io/pilaster/}), a collection of citation metadata extracted from publications in visualization for the digital humanities. The collection is generated from a seed set of relevant publications from which we extracted cited works, including journal and conference papers, books, theses, or blog posts, among other resources. The main aim of this work revolves around three main points: first, the collection may serve as an \textit{entry point} to the discipline for digital humanists and visualization scholars without previous experience in the field. Second, Pilaster can be regarded as a \textit{meeting point} for more established visualization or humanities scholars seeking to collaborate in the development of novel research ideas and related visualization design studies in the context of the humanities. Third, and given the large amount of visualization design spaces that were captured, we believe the dataset has the potential to become the \textit{starting point} for future studies aimed at understanding the particularities of problem-driven visualization research in this and other contexts.
} % end of abstract

\begin{document}

\maketitle

\section{Introduction}
The collaboration between computer scientists and humanities scholars presents a highly interesting field of experimentation that has produced important learning outcomes in the past and continues to do so until today. In general, and as it has occurred in other disciplines of science, applying computational methods to humanities research workflows has helped accelerate knowledge discovery and enhance the overall quality of results in humanistic research. A significant part of these combined efforts has typically focused on the application of data visualization techniques aimed at leveraging the interaction between humanities scholars and the said computational methods, producing interesting results in different conventional areas of humanistic research, such as discourse \cite{el-assady_visargue_2016}, literary \cite{hinrichs_speculative_2016}, or poetry analysis\cite{abdulrahman_rule-based_2013, mccurdy_poemage_2016, janicke_alignment_2018-1} or the browsing and sensemaking of cultural collections \cite{windhager_visualization_2018,windhager_exhibiting_2019}. 

However, the building and organization of interdisciplinary teams of experts that can produce valuable research outcomes in both the visualization and humanities domains seldom are problem-free \cite{simon_bridging_2015, janicke_valuable_2016}. Thus, this calls for special considerations to be taken into account by all the involved parties\cite{MiScKrLeKeEl+19}. This is particularly the case for visualization researchers new to the field whose previous experience may lie in other areas of visualization practice, and who may rapidly become overwhelmed by the complexities of the collaboration. Analogously, humanities scholars without previous or little experience in participating in visualization design studies may also encounter problems when trying to specify requirements and tasks due to their lack of visualization literacy \cite{pereda_tangible_2017}. 

The results presented in this paper constitute an extension of our recent work in the field\cite{8766090, benito-santos_data-driven_2020} that aims at supporting the immersion process\cite{hall_design_2020} of interdisciplinary researchers in visualization design studies within a digital humanities context, among other goals that are described throughout the paper. To this end, we employ a metadata collection of works on visualization for the digital humanities 
that we started building in 2019 and that we have kept curating and refining since then. The resulting dataset comprises almost 2,000 resources related to the practice of visualization in the context of digital humanities derived from an extensive analysis of the citations in a core set of 119 papers published at three different venues identified at the beginning of the study. In the following sections, we discuss the rationale we followed to build the dataset, and some of the problems we found in the process and which we could not fit into our previous contribution\cite{benito-santos_data-driven_2020} due to space limitations. Later, we present a description of the data fields and provide several descriptive statistics derived from the data that also offer new insight into the collection. Finally, we exemplify potential applications of \textit{Pilaster} with two simple use cases that others may find useful for carrying their own studies on the dataset. The first use case aims to capture our latest work on normalizing publication aggregation names, an effort that yielded new interesting insights into the commonalities and differences in venues commonly cited by DH and VIS researchers. In a second use case, we shed new light on how collaborations in the field are articulated, which suggests a lack of overlap between the two communities. 

\section{Surveying VIS4DH}
\label{sec:surveying}
\setlength{\intextsep}{2pt}%
\setlength{\columnsep}{4pt}%
\begin{wrapfigure}{l}{0.5\columnwidth} 
\fbox{\begin{minipage}{\dimexpr0.5\columnwidth-2\fboxrule-2\fboxsep}
    {\fontfamily{pbk}\selectfont \small
    "Let’s be honest—there is no definition of digital humanities, if by definition we mean a consistent set of theoretical concerns and research methods that might be aligned with a given discipline [...] How else to characterize the meaning of an expression that has nearly as many definitions as affiliates? It is a social category, not an ontological one."
    \vspace{5pt}
    
    R.C. Alvarado in \textit{The Digital Humanities Situation} (2011)\cite{alvarado_digital_2012}
    }
\end{minipage}}
\end{wrapfigure}

As mentioned in the previous section, the dataset arises from citations found in a core set of publications on visualization for the digital humanities. According to established literature review methodologies \cite{ed.20191026}, the process of literature review starts by defining the scope of the study, (\textit{"visualization for the digital humanities"} in our case). In the next step, the scope is condensed in a series of textual queries that are launched against online literature databases to obtain relevant publications. Then, these publications are analyzed, summarized, and discussed according to the classification dimensions and other traits derived by the authors of the survey \cite{ed.20191026}. Finally, the results of the review are wrapped up and prepared for dissemination to the scientific community. Whereas the process is seemingly straightforward, and we knew of similar methodologies that had been successfully applied to conduct surveys on specific sub-fields of the DH visualization practice \cite{windhager_visualization_2018}, it presents several issues that rendered it unfit for our purpose of capturing the different DH areas in which the visualization practice \textit{mostly} occurs. Besides, much of the work in digital humanities is presented exclusively at annual conferences (although some notable journals exist) whose proceedings are not indexed in the main online scientific databases. If, as it was our initial intention, our work should be aimed at interdisciplinary visualization practitioners, completely excluding all these works from an initial analysis seemed clearly counterproductive. Still, and beyond these considerations, we had to provide a sensible definition of the digital humanities to commence the survey. Here, we were facing a recurrent problem of the digital humanities that has been at the center of many academic debates. we resorted to the literature looking for a working definition of digital humanities that we could put to use, but we could not find any. \textit{How were we supposed to survey a topic that cannot be defined?}\cite{gold2012day}

As some authors like Alvarado have pointed out, the answer for the question of what the digital humanities are cannot rely on conventional conceptions of what a discipline should be\cite{alvarado_digital_2012}. Rather, he claims, it is more useful to see the digital humanities as a \textit{social category} that relates a collective of researchers who are involved in different, probably distant disciplines, and who call themselves "digital humanists." This statement was the cornerstone on which the methodology we adopted to generate the collection was built, and allowed us to move on to the data collection stage without the need to provide a definition of the digital humanities that would have stood on very shaky epistemological grounds, let alone a query that translated this definition into something that could be understood by a search engine. In such circumstances, we decided to adopt an utilitarian stance that focused instead on identifying the group of scholars who call themselves digital humanists \textit{and} practice visualization. Taking this reasoning further forward, it seemed obvious that this group must be composed of visualization practitioners interested in digital humanities, and also of digital humanists who have shown an interest for visualization. As we discuss in the next section, we looked for specific academic collectives whose members matched any of these two conditions. 

\section{Data Collection}
\label{sec:datacollection}
In this section, we detail how we built a seed dataset of publications from which the citations were extracted at a later stage. The methodology that we followed to build the dataset is inspired by other recent works in visualization research \cite{isenberg_vispubdata.org_2017,ed.20191026} that were adapted to cope with the diffuse character of digital humanities, as we explain in \autoref{sec:surveying}. 

\subsection{Sampling VIS authors}
\setlength{\intextsep}{0pt}%
\setlength{\columnsep}{7.5pt}%
\begin{wrapfigure}[30]{r}{0.45\columnwidth} 
    \vspace*{-5pt}
    \centering
    \includegraphics[width=0.45\columnwidth]{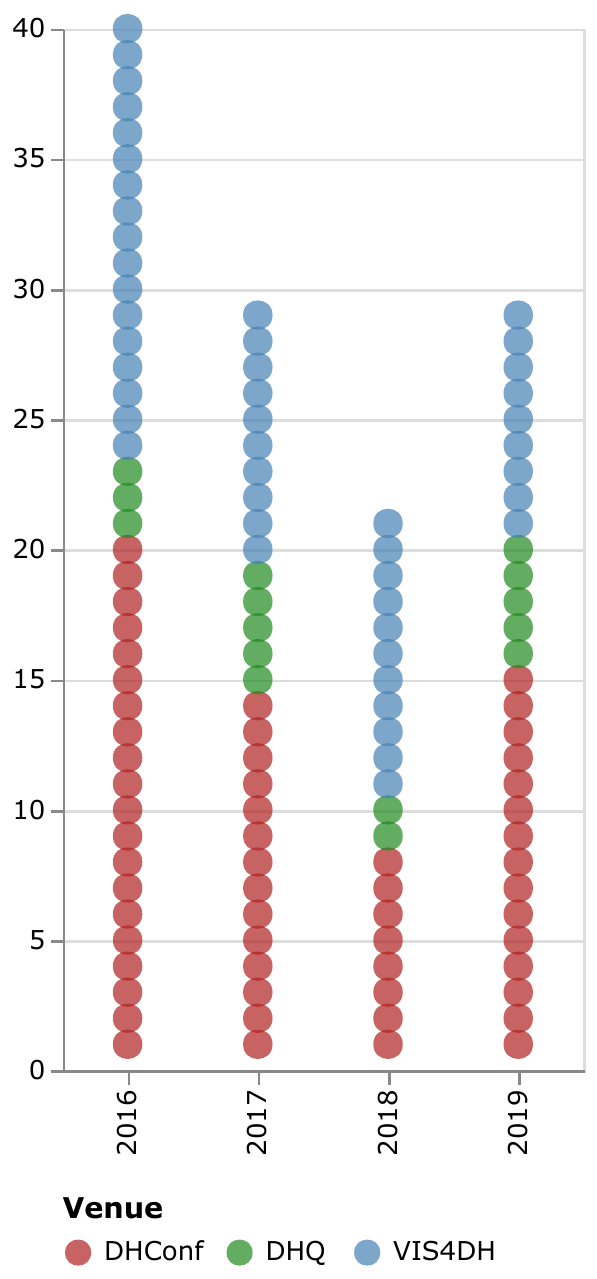}
  \caption{\label{fig:seed_by_year_and_venue} Dot plot showing the distribution by year and publication venue of papers in the seed dataset. A total of 119 papers were analyzed in a first stage.}
  \vspace*{-12.5pt}
\end{wrapfigure}

As explained before, the construction of the seed dataset involved the sampling of publications in both ends of the humanities-visualization collaboration. To find the components of the first group, we considered participants in the last editions of the VIS4DH workshop which is, to the best of our knowledge, the only space devoted to the task of "bringing together researchers and practitioners from the fields of visualization and the humanities to discuss new research directions at the intersection of visualization and (digital) humanities research." Although we knew of more research papers published at visualization conferences that could probably have been included in the seed dataset, we decided not do so due to the aforementioned impossibility of establishing a well-defined boundary between what qualifies as digital humanities and what not. At any rate, we assumed relevant papers would eventually appear during the analysis of the citations and therefore we preferred to keep the seed dataset as well scoped as possible. The inspection of the proceedings of the four first editions (2016-2019) of the workshop left a balance of 47 papers and 136 authors making up the first sample to be part of the seed dataset. 

\subsection{Sampling DH authors}
To obtain representative publications in the humanities side, we decided to inspect the proceedings of the last 4 editions (2016-2019) of the joint annual conference of the Alliance of Digital Humanities Organizations (ADHO) and its peer journal Digital Humanities Quarterly (DHQ). However, and unlike the previous case, we could not find a similar event to the VIS4DH in the DH Conference, and rather visualization practice seems to be spread across different areas such as geohumanities, linked open data, or audiovisuals \footnote{https://adho.org/special-interest-groups-sigs}. Given that we wanted to capture all works that employed visualization techniques regardless of their area of application, we opted for capturing long presentations and papers in the two venues that were related to visualization as tagged by their own authors. Concretely, we captured publications whose title, user-authored keywords or list topics (topics are chosen by the authors from a list of keywords compiled by ADHO) matched the regular expression “[Vv]isua*”. The search yielded a total of 72 publications (57 long presentations from the conference proceedings and 15 long papers from the journal) which constituted the "humanities" part of the seed dataset. The final composition of the seed dataset is shown in \autoref{fig:seed_by_year_and_venue}.

\section{Data Processing}
Publications in the first group were downloaded in PDF format from the workshop's homepage and their respective reference lists extracted with the pdftotext\footnote{https://github.com/jalan/pdftotext} library and stored for later processing. Reference lists of the second group were obtained by parsing the TEI-XML files in which the documents were encoded. The TEI files of publications in the DH Conference proceedings and the DHQ journal were obtained from the ADHO's GitHub repository \footnote{https://github.com/ADHO} and from the journal's website, respectively. The TEI files of the 2019 edition of the DH Conference had to be directly scraped from the conference website as they were missing from the repository. The bibliography sections of each paper in the seed dataset were analyzed with the Neural-ParsCit suite\cite{animesh2018neuralparscit}, which automatically extracts diverse metadata from text lines in a paper's reference list. The metadata includes but it is not limited to the title, publication year and venue, authors list, DOI and URL. For each of the extracted works, we completed their metadata with information obtained from the Elsevier API \footnote{https://dev.elsevier.com/} by matching their name with existing records in the database. Finally, author names and publication venues were normalized by following a semi-supervised iterative procedure that consisted in visually inspecting pairs displaying short edit distances. Whenever the names were found to refer to the same entity (author or venue), they were unified under their most common form. This process was repeated until no similar pairs were left. At the end of the extraction process, we obtained 2238 references of works that were cited from the seed dataset. They were resolved to 1934 different works of which 23 were publications originally included in the seed dataset. 

\section{Dataset Description}
\begin{figure}
\centering
\includegraphics[width=\columnwidth]{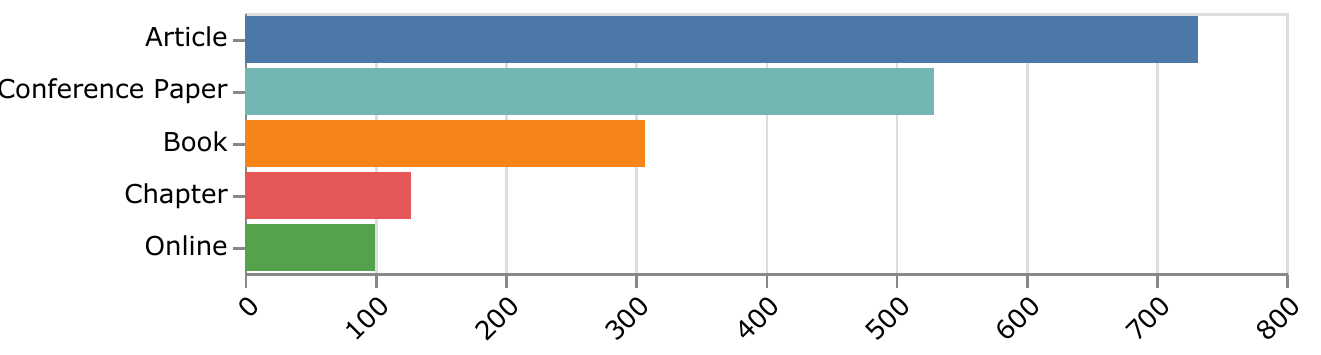}
\caption{Five most common resource types in the citations dataset. A majority of the cited items (1,263, 65.30\%) belong to one of the two top categories, although there are also references to books (307), book chapters (128) or online resources (100), such as blog posts or datasets.}
\label{fig:cites_by_type_ranked} 
\end{figure}

In this section, we describe the data fields that compose the entries and provide general descriptive statistics of the values they take. Below, we list data attributes that are common to items found in the seed or citations datasets:

\begin{itemize}[noitemsep]
    \item \textbf{key:} An automatically generated random key that identifies a given resource.
    \item \textbf{title:} The resource title obtained
    \item \textbf{authors:} The item's list of authors separated by semicolons. The complete list of authors comprises 3499 names of which 185 can also be found in the seed dataset. 
    \item \textbf{aggregation:} The normalized name of the aggregation in which the item can be found (e.g., a conference names or journal/book titles). We identified 1148 different aggregation types holding 1783 items in the citations dataset. 
    \item \textbf{year:} The year in which the item was created. It was obtained by parsing the reference or from the Elsevier API. 
    \item \textbf{source\_theme:} Denotes the provenance of the record. For items in the seed dataset, this field takes two values ("visualization", "humanities") depending on the type of the sample that included them, as described in \autoref{sec:datacollection}. The value is inherited by items in the citations dataset to annotate their provenance. Items cited from both parts of the seed dataset have this value set to "both". 
\end{itemize}
\begingroup
\setlength{\parindent}{0em}
Additionally, items in the \textbf{seed} dataset contain the following three extra fields:
\endgroup
\begin{itemize}[noitemsep]
    \item \textbf{publication\_short\_title:} An abbreviated form of publication\_title.  
    \item \textbf{author\_keywords:} Keywords list given by the items' authors.
    \item \textbf{n\_references:} Length of the reference list that can be found at the end of the paper. 
\end{itemize}
\begingroup
\setlength{\parindent}{0em}
Finally, data attributes exclusive to items in the \textbf{citations} dataset are listed below:
\endgroup
\begin{itemize}[noitemsep]
    \item \textbf{cited\_by:} A list of foreign keys pointing to papers in the seed dataset that cite the item. 
    \item \textbf{cited\_by\_venue:} Venue (VIS4DH, DH Conference, DHQ) of the paper(s) citing the item.
    \item \textbf{cited\_by\_count:} Number of papers in the seed dataset that cite a given item excluding self-references. We considered a citation to be a self-reference when the set intersection between the authors of the citing work and the authors of the cited work was not the empty set. 
    \item \textbf{type:} In cases where the publication could not be matched again an Elsevier record, we derived its type (e.g., conference paper, journal article, book) from other publications in the same venue that could be found. In total we identified 20 different cited work types (\autoref{fig:cites_by_type_ranked}).
    \item \textbf{aggregation\_type:} The type of the aggregation, if existent, in which the item can be found (e.g., journal, conference proceeding, or book).
    \item \textbf{link:} Web links extracted from the original reference that were parsed by means of a regular expression.
\end{itemize}
\begingroup
\setlength{\parindent}{0em}
In \autoref{fig:temp-source-whiskers}, we present some descriptive statistics that give an idea of the composition of the citations dataset according to its different dimensions. The first chart on the left shows how both communities follow similar temporal citation patterns with similar mean (2006) and median (2011) values. The next chart shows how the cited resources can be divided into three groups according to the community their citing counterparts belong to. As it can be seen in the figure, we obtained 280 resources that were referenced from VIS4DH and DHConference/DHQ papers, which in turn are among the most cited in the dataset: 82 out of the 100 most cited works belong to this category, which were cited a total of 267 times (11.93\% of all citations by papers in the seed dataset). Publications in this category are highly relevant because they represent the intersection point between the visualization and DH communities and therefore, 
they describe a shared communication channel \cite{simon_bridging_2015} between visualization and domain experts in DH research that we believe it is worth studying in greater depth. The seed and citations dataset were stored in a public spreadsheet (located at  \url{https://docs.google.com/spreadsheets/d/1Z8aMhxpai510hkuSVAFW6L4QyQfPPvUnv8IjKuF2_Jo/}) for ease of use by other researchers.

\endgroup

\section{Use Cases}
\label{sec:usecases}
In this section we propose two simple use cases of the dataset that can help to illustrate the potential use cases for the dataset. The first use case employs the citations dataset to explore commonly cited venues. Besides, we show other venues that are cited exclusively by researchers in one of the two sides. The second use case provides some insights on how interdisciplinary teams are conformed and how the collaborations are organized. The figures in this and the other sections of the paper were generated in Python code \footnote{\url{https://colab.research.google.com/drive/15cNprIDXsN1WMa660lo-ApimMib8vdth}} using the Vega-Lite grammar\cite{satyanarayan_vega-lite_2017} and Altair\cite{vanderplas_altair_2018}. 

\subsection{Studying publication aggregations}
\label{sec:pub-aggregations}
\begin{figure} 
    \centering
  \subfloat[]{%
       \includegraphics[width=1\columnwidth]{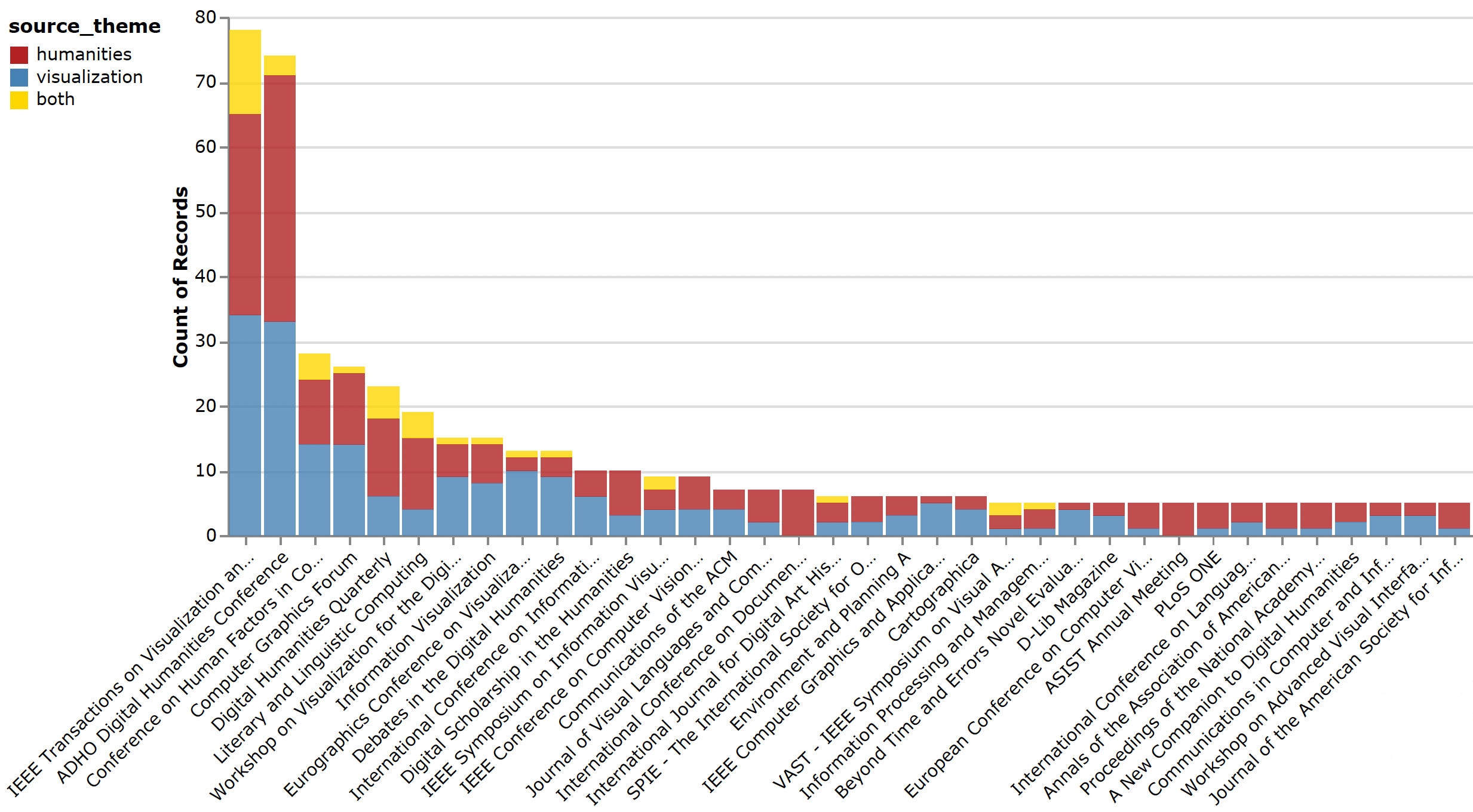}}
    \hfill
    \\
  \subfloat[]{%
        \includegraphics[width=1\columnwidth]{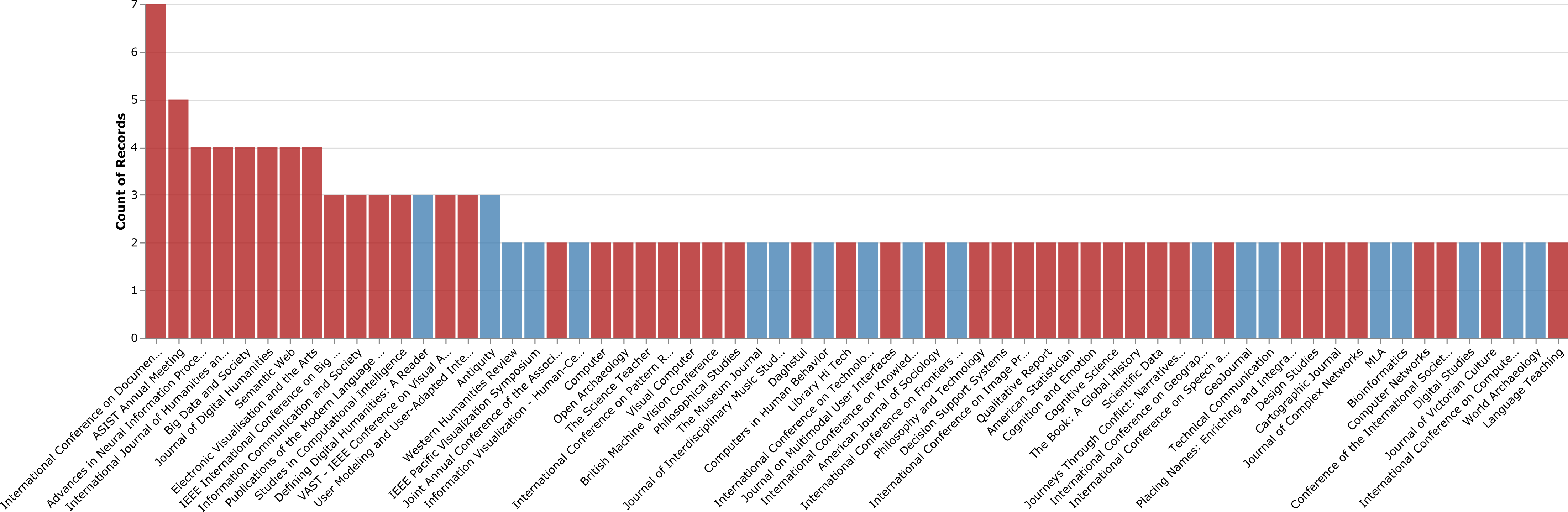}}
  \caption{\label{fig:exclusive-venues} (a) Top 5\% most cited aggregations in the citations dataset. IEEE Transactions on Visualization and Computer Graphics and the ADHO DH Conference are the most cited. (b) Most popular aggregations that are referenced exclusively by at least two different works in the VIS or DH seed datasets. }
\end{figure}

In this first use case, we are interested in exploring what venues are cited most often from what kinds of sources in the seed dataset. The stacked bar chart in Fig.\ref{fig:exclusive-venues}.a shows aggregations above the 95th percentile by number of times cited. From this visualization, some information can be decoded: for example, the two tallest bars in the chart depict the top two most cited venues, which are \textit{IEEE TVCG} and the \textit{DH Conference proceedings}. Moving to the right of the chart, other venues typically associated with visualization research appear, such as the \textit{Conference on Human Factors in Computing} (CHI), \textit{Computer Graphics Forum}, and the VIS4DH workshop, all of which are cited more or less evenly from the two categories of the seed dataset. A similar effect happens with other venues typical of DH research, among which we can find \textit{Digital Humanities Quarterly}, \textit{Digital Scholarship in the Humanities} and its previous title, \textit{Literary and Linguistic Computing}. As opposed to VIS venues, there seems to be a larger imbalance between the categories of items citing DH venues, which are mostly from works in the DH seed dataset. Closer to the tail of the distribution, we can detect other special venues that are \textit{exclusively} cited by publications originating in the DH domain, such as the \textit{International Conference on Document Analysis and Recognition} (ICDAR) or the \textit{Annual Meeting of the Association for Information Science and Technology} (ASIS\&T). We capture this idea in more detail in the chart of Fig.\ref{fig:exclusive-venues}.b, which represents venues cited exclusively by at least two publications in one of the two domains. These venues, we argue, may be indicative of current knowledge gaps in both sides of the visualization practice that could point to potential new areas for collaboration.

\subsection{Exploring the authors graph}
\label{sec:authors-graph}
\begin{figure}
\centering
\includegraphics[width=\columnwidth]{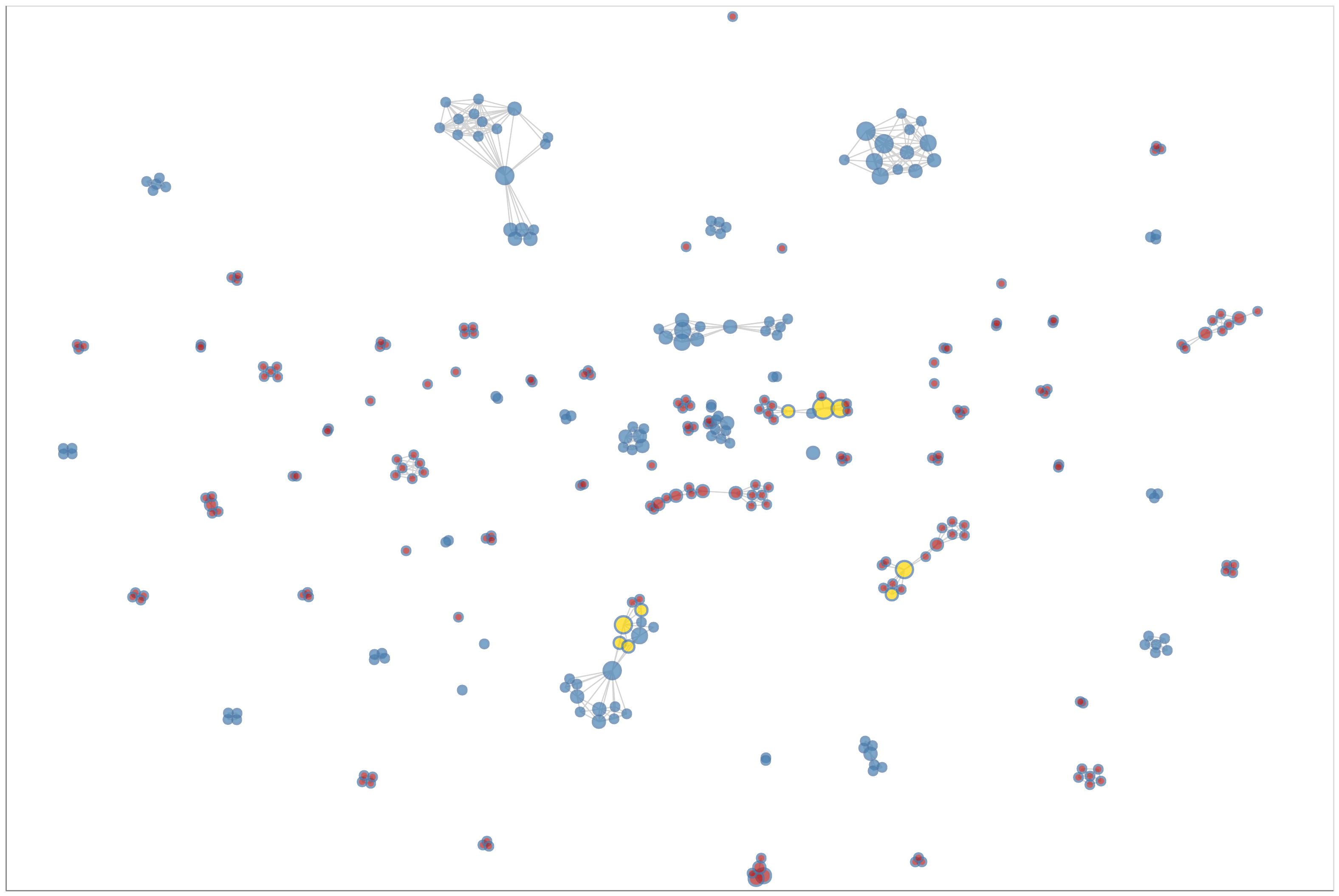}
\caption{A node-link diagram depicting co-authorship relationships between authors in the seed dataset. Only 9 out of 328 (2.76\%, in yellow) of individuals authored publications in both the VIS and the DH datasets.}
\label{fig:authors-graph}
\vspace{-5pt}
\end{figure}

In this second use case, we obtain insight into the size and structure of collaborations by means of a social network analysis of co-authorship relationships found in the seed dataset. The node-link diagram of \autoref{fig:authors-graph} depicts collaborations in both areas. By looking at the color of nodes in the chart, it can be seen that the number of authors who published papers in both categories is fairly (2.76\%) low, meaning that interactions between the two communities still are scarce, a fact that may be linked to certain issues pointed by other authors in the past\cite{janicke_valuable_2016}. Attending to the topology of the graph, author communities in the VIS side appear to be larger than their counterparts in the DH side, which may be partially due to differences in the average number of authors per paper in the two groups (DH: 2.92±1.51 vs VIS 4.15±2.10) but probably also to other factors that may deserve further study.

\section{Limitations and Future Work}
\subsection{Data Collection Methodology}
In \autoref{sec:datacollection}, we described the rationale that we followed to sample publications in both sides of the collaboration. Although we made an explicit effort to obtain a set of publications that was representative of the discipline, we are aware that, due to certain characteristics of the employed methodology, we might have missed previous work that could have been part of the seed dataset. For example, this could happen with VIS authors working on DH topics who have not participated in the VIS4DH workshop. A similar effect could happen with DH practitioners who decided not to include any terms matching the regular expression “[Vv]isua*” in their abstracts. In this respect, we expect to receive suggestions from the community of potential new sources that can be included as part of the collection in future developments to make it more complete. 

\subsection{Differences in Publication Formats}
The distribution of citations according to their provenance is skewed towards the humanities side, a phenomenon that can be traced to differences between the publication formats typically used on each domain. For example, whereas long presentations at the DH Conference are submitted as abstracts of maximum 750-1000 words, submissions to the VIS4DH workshop adopt the short paper form of 4+1 pages, which usually yield around 3500-4000 words ($\approx 3x$ longer). Although this difference in length is not translated into a similar difference in the average number of citations per paper between the two categories (right of \autoref{fig:temp-source-whiskers}), humanities papers consistently generated less citations on average than their VIS counterparts. However, they represent a thematically richer set of publications. Although we believe this fact is just representative of the reality of the field and it is not a drawback in itself, it is important to take it into account before extracting any conclusions from the dataset.

\subsection{Head or Tails}
In this paper, we tried to provide an overview of the collection by focusing on the heads of the rank-frequency distributions of, for example, resource types (\autoref{fig:cites_by_type_ranked}) or publication aggregations (\autoref{fig:exclusive-venues}). Whereas we believe this kind of analysis serves well the objective of describing the dataset, we are aware that this practice may also have unintended side effects: for example, it could happen that these rank-frequency distributions may be interpreted as importance rankings that go beyond the purpose of providing an entry point to the dataset, a practice which we have argued against in the past\cite{8766090}. By looking \textit{only} at top-ranked items while disregarding the rest, other vital information for advancing the field may be missed, a practice that also dangerously contributes toward perpetuating prestige bias (among other biases) in academia. Rather, we recommend potential users of the collection to repair on items found at the tails of the distributions, for example by performing searches on specific terms   that could unveil highly-interesting but lowly-cited, underrepresented themes, works, venues, or authors. 

\section{Conclusion}
In this paper, we presented \textit{Pilaster}, a metadata collection of papers and related citations employed by scholars working at the intersection of visualization and digital humanities. By departing from a representative sample of publications in the field, we aimed at capturing the different perspectives of scholars at both ends of the collaboration. Furthermore, we exemplified how insight into the discipline can be obtained by means of two use cases that can be easily adapted by other researchers to cover more complex interactions and usage scenarios. In addition, the resulting spreadsheet and code used to generate the figures in this paper were put in the public domain and can be consulted online. Beyond serving as an entry point to the discipline for novel researchers to the field, the results of our work are also aimed at more established scholars who may find them useful for detecting potential future collaborations or novel research ideas, as we illustrated in \autoref{sec:usecases}. Although we plan to continue updating the dataset as new publications become available, we encourage other researchers to send us feedback or suggestions of other use cases that we may not have covered here. 

\bibliographystyle{abbrv-doi}

\bibliography{template}
\end{document}